\documentclass[letterpaper, 10 pt, conference]{ieeeconf}  

\IEEEoverridecommandlockouts                              

\usepackage{amsfonts}
\usepackage{xcolor}
\usepackage{graphicx} 
\usepackage{epsfig} 
\usepackage{cite}

\usepackage{subcaption}
\usepackage{stfloats}
\usepackage{amsmath} 
\usepackage{siunitx}
\usepackage{multicol}

\makeatletter

\newcommand{\Rmnum}[1]{\expandafter\@slowromancap\romannumeral #1@}
\makeatother
\DeclareMathOperator*{\argmax}{arg\,max}

\usepackage{booktabs}
\newcommand{\ra}[1]{\renewcommand{\arraystretch}{#1}}
\usepackage{mathtools}
\newtheorem{remark}{Remark}
\usepackage{cite}
\usepackage{flushend}
\usepackage{algorithm}
\usepackage{algpseudocode}
\usepackage{empheq}

\definecolor{darkred}{rgb}{0.6, 0, 0}
\definecolor{darkblue}{rgb}{0, 0.4, 1}
\def\p(#1|#2){p(#1\,|\,#2)}

\title{\LARGE \bf
System Identification for Lithium-Ion Batteries with Nonlinear Coupled Electro-Thermal Dynamics via Bayesian Optimization
}

\author{Hao Tu{$^1$}, Xinfan Lin{$^2$} , Yebin Wang{$^3$}, Huazhen Fang{$^1$}
\thanks{{$^1$}Hao Tu and Huazhen Fang are with the Department of Mechanical Engineering, University of Kansas, Lawrence, KS 66045, USA.
        {\tt\small tuhao@ku.edu, fang@ku.edu}}%
\thanks{{$^2$}Xinfan Lin is with the Department of Mechanical and Aerospace Engineering, University of California, Davis, CA 95616, USA.
        {\tt\small lxflin@ucdavis.edu}}%
\thanks{{$^3$}Yebin Wang is with the Mitsubishi Electric Research Laboratories, Cambridge, MA 02139, USA.
        {\tt\small yebinwang@ieee.org}}%
}

\begin{document}

\maketitle
\thispagestyle{empty}
\pagestyle{empty}

\begin{abstract}

Essential to various practical applications of lithium-ion batteries is the availability of accurate equivalent circuit models. This paper presents a new coupled electro-thermal model for batteries and studies how to extract it from data. We consider the problem of maximum likelihood parameter estimation, which, however, is nontrivial to solve as the model is nonlinear in both its dynamics and measurement. We propose to leverage the Bayesian optimization approach, owing to its machine learning-driven capability in handling complex optimization problems and searching for global optima. To enhance the parameter search efficiency, we dynamically narrow and refine the search space in Bayesian optimization. The proposed system identification approach can efficiently determine the parameters of the coupled electro-thermal model. It is amenable to practical implementation, with few requirements on the experiment, data types, and optimization setups, and well applicable to many other battery models.
\end{abstract}

\section{Introduction}
The global push for sustainability and decarbonization has driven the widespread use of lithium-ion batteries (LiBs) as power sources or energy storage systems for electric mobility, smart grid, and renewable energy. In all these applications, battery management systems run to ensure the safety, performance, and longevity of LiBs from cell to system level. They often adopt equivalent circuit models (ECMs) to predict the behaviors of LiBs, because ECMs can strike a desired balance between predictive accuracy and computational efficiency~\cite{Hu:JPS:2012}.

The existing body of literature has introduced a range of ECMs. After choosing an ECM for LiBs, it becomes crucial to determine the model parameters. A popular methodology among practitioners for this purpose is experiment-based parameter calibration. Its core idea is that one can design and implement specific current profiles on a LiB cell to stimulate relevant dynamic processes and excite the effects of concerned parameters on the measurements. Various studies about the Thevenin model leverage voltage transient responses under pulse current charging/discharging to determine the internal resistance and capacitance parameters~\cite{Chen:TEC:2006,Fairweather:JPS:2011}. It is also a common practice to use trickle constant current charging/discharging to calibrate the mapping from the state-of-charge (SoC) to the open-circuit voltage (OCV)~\cite{Dubarry:JPS:2007}. Some studies have proposed specialized charging/discharging protocols that involve systematized testing procedures to extract a complete ECM~\cite{Samieian:Batt:2022,Biju:AE:2023}. Experimental calibration techniques are easy to comprehend and handy to implement, but they provide barely sufficient accuracy, restrict to some specific current profiles, and require long hours of testing. 

System identification represents a more formalized and mathematical approach to extracting ECMs from data. In this regard, a widely adopted method is to minimize the prediction error of an ECM with respect to the measurements. This generally leads to the formulation of nonlinear optimization problems. For example, nonlinear least squares are considered in~\cite{Sitterly:ITSE:2011,Fend:JPS:2015,Tian:JES:2020} to perform parameter estimation for ECMs. Some recent studies deal with system identification for ECMs from the viewpoint of statistical inference. They pose maximum likelihood or maximum a posteriori estimation problems~\cite{Tian:TCST:2021} and then recast them as optimization problems to solve for the unknown parameters. In general, optimization problems for ECM identification are tractable under some strong conditions like constant current excitation and linear dynamics in the ECM. However, they can become highly nonlinear and nonconvex in other cases, e.g., when one uses arbitrary or variable current profiles, or when the ECM is nonlinear in its dynamics.  Gradient-based optimization will easily get stuck in local optima and be sensitive, even vulnerable, to initial parameter guesses. Global optimization methods, e.g., particle swarm optimization~\cite{Hu:JPS:2012,Yu:TIE:2017} and the Cuckoo search~\cite{Duru:JES:2022,Li:ESM:2022}, thus gain use in some studies to identify ECMs, but at the expense of high computational complexity.

When ECMs capture coupled electro-thermal dynamics, they present more challenges for system identification. This is because the interaction between the electrical and thermal submodels will result in more unknown parameters, stronger nonlinearity, and lower parameter identifiability. For such ECMs, it is possible to parameterize the temperature dependence and then calibrate the ECMs at different temperatures~\cite{Chen:TEC:2006}. An alternative way is to control the testing conditions so that only either the electrical or the thermal dynamics will be excited and identified at a time~\cite{Perez:ASME:2012,Lin:TCST:2013,Samad:DSMC:2017}. These techniques, however, either require many experiments, or involve empiricism or approximations to limit the achievable level of accuracy. It thus remains a challenge to identify ECMs with coupled electro-thermal dynamics.

In this paper, we consider a new electro-thermal model for LiBs, which results from adding the temperature dependence to the nonlinear double capacitor (NDC) model in~\cite{Tian:TCST:2021}. Called NDC-T, the model has nonlinear dynamics and measurement, while involving a few more parameters than other ECMs of similar kind in the literature. To successfully extract it from data, we consider a maximum likelihood parameter estimation problem and solve it using Bayesian optimization (BayesOpt)~\cite{frazier:tutorial:2018}. BayesOpt is a machine learning approach for optimization---it builds a data-driven probabilistic surrogate model for the objective function, iteratively updates the surrogate, and uses the surrogate to search for optima by balancing exploitation (searching promising regions) and exploration (searching uncertain regions). This approach is capable of handling objective functions difficult to evaluate, obviating the use of gradients, and finding global optima in complex search spaces. Gaining from the benefits of BayesOpt, we can accurately identify the physical parameters of the continuous-time NDC-T model directly from measurement data based on a wide range of current profiles. We also introduce a procedure to dynamically shrink the parameter search space to accelerate the search and computation. The proposed system identification approach well lends itself to other ECMs and electrochemical models.

The remainder of the paper is organized as follows. Section~\ref{sec: Model} introduces the NDC-T model and the parameter identification problem. In Section~\ref{sec: ParamID}, we propose our BayesOpt-based system identification approach. Section~\ref{sec: Sim} provides a simulation study to validate the approach. Finally, Section~\ref{sec: Conclusion} ends the paper with conclusions.

\section{NDC-T Model and Parameter Identification Problem} \label{sec: Model}

This section presents the NDC-T model, a coupled electro-thermal model for LiBs, and the challenges in its parameter identification.

\begin{figure}[t!]
    \centering
    \includegraphics[width = .4\textwidth,trim={10.1cm 3.1cm 13.3cm 2.2cm},clip]{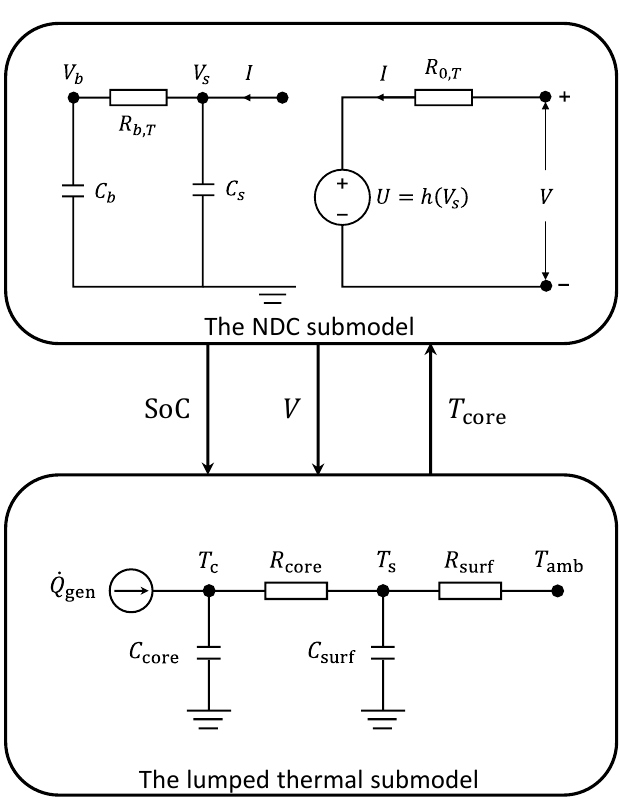}
    \caption{The NDC-T model, which couples the NDC submodel and the lumped thermal submodel.}
    \label{Fig: NDC-T model}
\end{figure}

The NDC-T model integrates the NDC model in~\cite{Tian:TCST:2021} and the lumped thermal model in~\cite{Lin:JPS:2014} to capture the electro-thermal behavior of LiBs, as shown in Fig.~\ref{Fig: NDC-T model}. The NDC model uses equivalent electrical circuits to approximate both the lithium-ion diffusion and nonlinear voltage behavior inside a LiB cell. The model uses an RC chain composed of $R_{b,T}$, $C_b$ and $C_s$ so that the charge transfer between $C_b$ and $C_s$ simulates the migration of lithium ions between the core and surface of an electrode. Note that $V_b$ and $V_s$, the respective voltage across the $C_b$ and $C_s$, are analogous to the lithium-ion concentrations at the core and surface, respectively, and that $R_{b,T}$ emulates the lithium-ion diffusion resistance~\cite{Fang:TCST:2017}.  Coupled with the RC chain is the open-circuit voltage (OCV) source $U = h_{\mathrm{OCV}}(V_s)$ and the internal resistor $R_{o,T}$. The governing equations of the NDC model are given by:
\begin{align*}
\begin{bmatrix}
\dot{V_b}(t) \\ \dot{V_s}(t)
\end{bmatrix}
&= 
\begin{bmatrix}
    \frac{-1}{C_b R_{b,T}} & \frac{1}{C_b R_{b,T}} \\
    \frac{1}{C_s R_{b,T}} & \frac{-1}{C_s R_{b,T}}
\end{bmatrix}
\begin{bmatrix}
V_b(t) \\ V_s(t)
\end{bmatrix} 
+ 
\begin{bmatrix}
    0\\
    \frac{1}{C_s}
    \end{bmatrix}
I(t),\\ 
V(t) &= h_{\mathrm{OCV}}(V_s(t)) + R_{o,T} I(t),
\end{align*}
where $I$ is the input current, with $I<0$ for discharging, and $I>0$ for charging.
The SoC is given by
\begin{align*} 
    \mathrm{SoC} = \frac{C_bV_b+C_sV_s}{C_b+C_s} \times 100\%.
\end{align*}
Here, we have $V_b=V_s=0$ V when $\mathrm{SoC} = 0\%$, and $V_b=V_s=1$ V when $\mathrm{SoC}=100\%$.

We consider a cylindrical LiB cell and use an equivalent thermal circuit to capture its thermal dynamics. As shown in Fig.~\ref{Fig: NDC-T model}, the circuit lumps the spatial temperature distribution into the temperatures at the core and the surface, which are denoted as $T_c$ and $T_s$. This lumped thermal model is given by
\begin{align*} 
\begin{bmatrix}
\dot{T}_c(t) \\ \dot{T}_s(t)
\end{bmatrix}
&= 
\begin{bmatrix}
    \frac{-1}{R_{\mathrm{core}} C_{\mathrm{core}}} & \frac{1}{R_{\mathrm{core}} C_{\mathrm{core}}} \\
    \frac{1}{R_{\mathrm{core}} C_{\mathrm{surf}}} & \frac{-1}{R_{\mathrm{surf}} C_{\mathrm{surf}}}+\frac{-1}{R_{\mathrm{core}} C_{\mathrm{surf}}}
    \end{bmatrix}
\begin{bmatrix}
T_c(t) \\ T_s(t)
\end{bmatrix} \\
& \quad + \begin{bmatrix}
    \frac{1}{C_{\mathrm{core}}} & 0\\
    0 & \frac{1}{R_{\mathrm{surf}} C_{\mathrm{surf}}}
    \end{bmatrix}
\begin{bmatrix}
    \dot{Q}_{\mathrm{gen}}(t) \\
    T_{\mathrm{amb}}(t)
\end{bmatrix}.
\end{align*}
Here, $T_{\mathrm{amb}}$ is the ambient temperature; $C_{\mathrm{core/surf}}$ is the heat capacity at the cell's core/surface; $R_\mathrm{core}$ is the conduction resistance between the cell's core and surface; $R_\mathrm{surf}$ is the convection resistance between the cell's surface and the environment; $\dot Q_\mathrm{gen}$ is the heat generation rate, which is assumed to be concentrated in the core of the cell.

Next, we join the two models so that they will interact with each other to describe the cell's dynamics with higher fidelity. The coupling involves two aspects. First, the internal resistance $R_{o,T}$ and the diffusion resistance $R_{b,T}$ are made dependent on $T_c$ following the Arrhenius law
\begin{align*}
    R_{o,T} &= R_o \cdot \exp{\left(\kappa_1 \left(\frac{1}{T_c}-\frac{1}{T_{\mathrm{ref}}}\right)\right)}, \\
    R_{b,T} &= R_b \cdot \exp{\left(\kappa_2 \left(\frac{1}{T_c}-\frac{1}{T_{\mathrm{ref}}}\right)\right)},
\end{align*}
where $\kappa_1$ and $\kappa_2$ are coefficients, and $T_{\mathrm{ref}}$ is the reference temperature. Second, the heat generation term $\dot Q_\mathrm{gen}$ is a function of the voltage and SoC
\begin{align*}
\dot{Q}_{\mathrm{gen}}=I(V-h_\mathrm{OCV}(\mathrm{SoC})).
\end{align*}

With the coupling, we obtain the NDC-T model that is able to characterize the electro-thermal dynamics of the cell. The model is supplemented with two measurements, i.e., the terminal voltage $y_V$ and surface temperature $y_T$. The measurements are subject to noises due to imperfect sensors:
\begin{align*}
    y_{V}(t) &= V(t) + e_V, \\
    y_{T}(t) &= T_s(t) + e_T,
\end{align*}
where $e_V$$\sim$$\mathcal{N}(0,R_V)$ and $e_T$$\sim$$\mathcal{N}(0,R_T)$ are white Gaussian noises.

For simplicity, we summarize and rewrite the NDC-T model compactly as follows:
\begin{subequations} \label{Eqn: summary model Total}
\begin{empheq}[left=\empheqlbrace]{align} \label{Eqn: summary model}
      &\dot{\boldsymbol{x}} = f(\boldsymbol{x},\boldsymbol{u}), \\
      &\boldsymbol{y} = h(\boldsymbol{x},\boldsymbol{u}) + \boldsymbol{e},
\end{empheq}
\end{subequations}
where $\boldsymbol{x} = [V_b\ V_s\ T_c\ T_s]^\top$, $\boldsymbol{u} = [I \ T_{\mathrm{amb}}]^\top$, $\boldsymbol{y} = [y_{V}\ y_{T}]^\top$, $\boldsymbol{e} = [e_V \  e_T]^\top$, $f$ is the nonlinear state function, and $h$ is the nonlinear output function. Based on~(\ref{Eqn: summary model Total}), it is of our interest to extract the NDC-T model parameters from measurement data.

We introduce the parameter identification for the NDC-T model from the perspective of maximum likelihood estimation. To formulate a tractable problem, we introduce two reasonable assumptions for the NDC-T model. First, the SoC-OCV function $h_\mathrm{OCV}(\cdot)$ has been determined beforehand using some techniques like trickle current charging/discharging~\cite{Plett:Vol2:2015}. Second the model has no process noise.
For the NDC-T model, the unknown parameters to identify include
\begin{align*}
    \boldsymbol{\theta} = 
    \begingroup
    \renewcommand*{\arraycolsep}{3pt}
    \begin{bmatrix} 
        C_b & C_s & R_b & R_o & C_{\mathrm{core}} & C_{\mathrm{surf}} & R_{\mathrm{core}} & R_{\mathrm{surf}} & \kappa_1 & \kappa_2
    \end{bmatrix}^\top, 
    \endgroup
\end{align*}
where $\boldsymbol{\theta}$ denotes the parameter vector. 

To estimate $\boldsymbol{\theta}$, we apply a current sequence to the LiB cell and collect the voltage and temperature measurements at consecutive time instants $t_1, \ldots, t_N$. The dataset thus includes
\begin{align*}
\boldsymbol{u}_{1:N} = \begin{bmatrix}
        \boldsymbol{u}_{1} & \cdots &  \boldsymbol{u}_{N}
    \end{bmatrix}^\top,
    \quad
\boldsymbol{y}_{1:N} = \begin{bmatrix}
        \boldsymbol{y}_{1} & \cdots &  \boldsymbol{y}_{N}
    \end{bmatrix}^\top,
\end{align*}
where $\boldsymbol{u}_n = \boldsymbol{u}(t_n)$, $\boldsymbol{y}_n = \boldsymbol{y}(t_n)$.
Our goal is to obtain $\hat{\boldsymbol{\theta}}$ which maximizes the log-likelihood of the measurements:
\begin{align} \label{Eqn: ML}
    \hat{\boldsymbol{\theta}} = \argmax_{\boldsymbol{\theta}} L(\boldsymbol{\theta}) \coloneqq \log \p(\boldsymbol{y}_{1:N}|\boldsymbol{u}_{1:N}, \boldsymbol{\theta}),
\end{align}
where $\p(\boldsymbol{y}_{1:N}| \boldsymbol{u}_{1:N}, \boldsymbol{\theta})$ is the likelihood distribution of the measurements conditioned on $\boldsymbol{u}_{1:N}$ and $\boldsymbol{\theta}$.

However, the system identification problem in~(\ref{Eqn: ML}) is nontrivial. The primary cause is that the model is nonlinear, presents itself in the state-space form, and involves a good number of physical parameters. This will give rise to several challenges. 

\begin{itemize}

\item The model is continuous-time, and discretization of it will introduce errors. Such errors will propagate into the identification to reduce the parameter estimation accuracy. 

\item The model's nonlinearity will lead to complex nonconvex optimization landscapes. Gradient-based optimization will struggle to find global optima. It could be a brittle solution here, because of the sensitivity to initial guesses and the difficulty to find or compute the gradients of the objective function. 

\item The large parameter space, due to the 10 unknown parameters, will hinder the performance and chance of success in parameter search.

\end{itemize} 

To tackle the challenges, we will develop a BayesOpt-based approach to treat the problem in~(\ref{Eqn: ML}).

\section{Parameter Identification Approach} \label{sec: ParamID}

This section proposes our solution to the NDC-T's identification problem in Section~\ref{sec: Model}. First, we present a Monte Carlo simulation approach to evaluate $L(\boldsymbol{\theta})$ at a given $\boldsymbol{\theta}$. Then, we briefly introduce the two components of BayesOpt. Finally, we present a search space reduction scheme for a more efficient parameter search.

\subsection{Sampling-Based Evaluation of $L(\boldsymbol{\theta})$}
The evaluation of $\p(\boldsymbol{y}_{1:N}|\boldsymbol{u}_{1:N}, \boldsymbol{\theta})$ is difficult, as there is no analytical solution because the NDC-T model is nonlinear. We thus turn to Monte Carlo sampling to perform numerical evaluation. Using Bayes' rule, we have 
\begin{align} \label{Eqn: yutheta}
\nonumber
    &\p(\boldsymbol{y}_{1:N}|\boldsymbol{u}_{1:N}, \boldsymbol{\theta}) \\ &= \int \p(\boldsymbol{y}_{1:N}|\boldsymbol{x}_{1:N},\boldsymbol{u}_{1:N}, \boldsymbol{\theta})\p(\boldsymbol{x}_{0:N}|\boldsymbol{u}_{1:N}, \boldsymbol{\theta}) \, d\boldsymbol{x}_{0:N},
\end{align}
where
$\boldsymbol{x}_{0:N} =
    \begin{bmatrix} 
        \boldsymbol{x}_0  & \cdots & \boldsymbol{x}_N
    \end{bmatrix}^\top$, $\boldsymbol{x}_n = \boldsymbol{x}(t_n)$.
Considering that $p(\boldsymbol{x}_0)$ is known based on the cell's initial condition, we can build an empirical distribution for $\p(\boldsymbol{x}_{0:N}|\boldsymbol{u}_{1:N},\boldsymbol{\theta})$:
\begin{align} \label{Eqn: Sampling}
\p(\boldsymbol{x}_{0:N}|\boldsymbol{u}_{1:N},\boldsymbol{\theta}) \approx \frac{1}{N_p} \sum_{i=1}^{N_p}  \delta(\boldsymbol{x}_{0:N}-\boldsymbol{x}_{0:N}^{(i)}),
\end{align}
where $\boldsymbol{x}_{0:N}^{(i)}$ for $i=1,2,\ldots,N_p$ is is the $i$-th sample-based state trajectory by running the NDC-T model forward. Given~(\ref{Eqn: Sampling}),~(\ref{Eqn: yutheta}) can be approximated as
\begin{align*}
    \nonumber
    \p(\boldsymbol{y}_{1:N}| \boldsymbol{u}_{1:N}, \boldsymbol{\theta}) &\approx \frac{1}{N_p} \sum _{i=1}^{N_p} \p(\boldsymbol{y}_{1:N}|\boldsymbol{x}_{1:N}^{(i)}, \boldsymbol{u}_{1:N}, \boldsymbol{\theta}) \\
    \nonumber
    &= \frac{1}{N_p} \sum _{i=1}^{N_p}  \prod_{n=1}^{N} \p(\boldsymbol{y}_{n}|\boldsymbol{x}_{n}^{(i)}, \boldsymbol{u}_{n}, \boldsymbol{\theta}).
\end{align*}
Therefore, $L(\boldsymbol{\theta})$ is approximated by
\begin{align}\label{Eqn: likelihood computation}
    L(\boldsymbol{\theta}) \approx -\log {N_p} + \log \sum _{i=1}^{N_p} \prod_{n=1}^{N} \p(\boldsymbol{y}_{n}|\boldsymbol{x}_{n}^{(i)}, \boldsymbol{u}_{n}, \boldsymbol{\theta}).
\end{align}
Note that sampling from $p(\boldsymbol{x}_{0:N}| \boldsymbol{u}_{1:N}, \boldsymbol{\theta})$ is hard for general nonlinear state space systems~\cite{Durbin:Bio:1997}. However, since~(\ref{Eqn: summary model}) of the NDC-T model is deterministic, we can compute $\boldsymbol{x}_{1:N}$ given $\boldsymbol{\theta}$ by running~(\ref{Eqn: summary model}) forward after sampling $\boldsymbol{x}_{0}$. The model run can be based on the Runge-Kutta method or other numerical methods. For high accuracy, we can use a small enough step size.


\begin{remark}
    The above maximum likelihood-based formulation for parameter estimation can be easily extended to the case of multiple datasets. Consider that $M$ datasets $\boldsymbol{y}^{1:M} = \left\{ \boldsymbol{y}_{1:N_1}^{1}, \cdots, \boldsymbol{y}_{1:N_M}^{M}\right\}$ and $\boldsymbol{u}^{1:M} = \left\{ \boldsymbol{u}_{1:N_1}^{1}, \cdots, \boldsymbol{u}_{1:N_M}^{M}\right\}$ are collected independently, we have
    \begin{align*}
        \log \p(\boldsymbol{y}^{1:M}|\boldsymbol{u}^{1:M},\boldsymbol{\theta}) = \sum_{m=1}^{M} \log \p(\boldsymbol{y}_{1:N_m}^{m}|\boldsymbol{u}_{1:N_m}^{m},\boldsymbol{\theta}).
    \end{align*}
    This implies that the total log-likelihood given a $\boldsymbol{\theta}$ will be the sum of the log-likelihoods for each dataset. Mostly, the use of more datasets will improve the accuracy in parameter estimation.
\end{remark}

\subsection{BayesOpt}




















BayesOpt considers $L(\boldsymbol{\theta})$ as a black-box function and uses Gaussian processes to capture probabilistic relations between $\boldsymbol{\theta}$ and $L(\boldsymbol{\theta})$. The Gaussian process will serve as a surrogate model to approximate $L(\boldsymbol{\theta})$ and be used to search for the optimal solution. For clarity, we denote the surrogate for $L(\boldsymbol{\theta})$ as $\hat{L}(\boldsymbol{\theta})$. 
As a Gaussian process, $\hat L(\boldsymbol{\theta})$ takes the following prior distribution:
\begin{align*}
    \hat{L}(\boldsymbol{\theta}) \sim \ \mathcal{GP}(\mu(\boldsymbol{\theta}), k(\boldsymbol{\theta}, \boldsymbol{\theta}')),
\end{align*}
where $\mu(\cdot)$ and $k(\cdot,\cdot)$ are the mean and kernel functions. Note that $k(\boldsymbol{\theta}, \boldsymbol{\theta}')$ encodes the correlation between $\hat{L}(\boldsymbol{\theta})$ and $\hat{L}(\boldsymbol{\theta}')$, where $\boldsymbol{\theta}$ and $\boldsymbol{\theta}'$ belong to the parameter space.  Common choices of $k(\cdot,\cdot)$ include the squared exponential kernel and M{\`a}tern kernel~\cite{frazier:tutorial:2018}. Given $q$ parameters and their corresponding log-likelihood values $L(\boldsymbol{\theta}_{1:q}) = \left[ L(\boldsymbol{\theta}_1)\ \cdots\ L(\boldsymbol{\theta}_q) \right]^\top$, the posterior distribution of $\hat{L}(\boldsymbol{\theta})$ can be obtained as
\begin{align} \label{Eqn: GP posterior}
    \hat{L}(\boldsymbol{\theta}) \; | \; L(\boldsymbol{\theta}_{1:q}) \sim \mathcal{N} (\mu_q(\boldsymbol{\theta}), \Sigma_q(\boldsymbol{\theta})),
\end{align}
where 
\begin{align*}
    \mu_q(\boldsymbol{\theta}) &= \mu(\boldsymbol{\theta}) + \bar{k}(\boldsymbol{\theta},\boldsymbol{\theta}_{1:q})K^{-1}\left(L(\boldsymbol{\theta}_{1:q})-\mu(\boldsymbol{\theta}_{1:q})\right), \\
    \Sigma_q(\boldsymbol{\theta}) &= k(\boldsymbol{\theta},\boldsymbol{\theta})-\bar{k}(\boldsymbol{\theta},\boldsymbol{\theta}_{1:q})K^{-1}\bar{k}(\boldsymbol{\theta})^\top.
\end{align*}
Here,
\begin{align*}
    \mu(\boldsymbol{\theta}_{1:q}) &= 
    \begin{bmatrix}
        \mu(\boldsymbol{\theta}_1)\ \cdots\ \mu(\boldsymbol{\theta}_q)
    \end{bmatrix}^\top,\\
    \bar{k}(\boldsymbol{\theta},\boldsymbol{\theta}_{1:q}) &= 
    \begin{bmatrix}
        k(\boldsymbol{\theta}, \boldsymbol{\theta}_1) & \cdots & k(\boldsymbol{\theta}, \boldsymbol{\theta}_q)
    \end{bmatrix}, \\
    K &=  
    \begin{bmatrix}
         k(\boldsymbol{\theta}_1, \boldsymbol{\theta}_1) & \cdots & k(\boldsymbol{\theta}_1, \boldsymbol{\theta}_q) \\
         \vdots & \ddots & \vdots \\
         k(\boldsymbol{\theta}_1, \boldsymbol{\theta}_q)^\top & \cdots & k(\boldsymbol{\theta}_q, \boldsymbol{\theta}_q)
    \end{bmatrix}.
\end{align*}
The posterior distribution in~(\ref{Eqn: GP posterior}) represents the prediction of $\hat{L}(\boldsymbol{\theta})$ based on the existing data points. It will be used to find out the next sample $\boldsymbol{\theta}_{q+1}$.

BayesOpt uses the so-called acquisition function to guide the search for $\boldsymbol{\theta}_{q+1}$. For the acquisition function design, a popular choice is the expected improvement. The improvement refers to the increase of $\hat{L}(\boldsymbol{\theta})$ with respect to the maximum of the so far observed $L(\boldsymbol{\theta}_{1:q})$. As $\hat{L}(\boldsymbol{\theta})$ is probabilistic, we must consider the expectation of the improvement. Specifically, denoting $L^* = \max L(\boldsymbol{\theta}_{1:q})$, the expected improvement is defined as
\begin{align*}
    \mathrm{EI(\boldsymbol{\theta})} = \mathbb{E} \Big[ \Big( \hat{L}({\boldsymbol{\theta}}) - L^* \Big)^+ \; \Big| \; L(\boldsymbol{\theta}_{1:q}) \Big],
\end{align*}
where $(\cdot)^+ = \max(\cdot,0)$, and the expectation is taken over the posterior distribution given by~(\ref{Eqn: GP posterior}). Then, $\boldsymbol{\theta}_{q+1}$ is selected to be the point that maximizes  $\mathrm{EI(\boldsymbol{\theta})}$, that is
\begin{align*}
    \boldsymbol{\theta}_{q+1} = \argmax_{\boldsymbol{\theta}} \mathrm{EI(\boldsymbol{\theta})}.
\end{align*}
The expected improvement-based search for $\boldsymbol{\theta}_{1:q}$ will not only exploit the available knowledge, embodied by $L(\boldsymbol{\theta}_{1:q})$, but also explore the parameter space by harnessing the probabilistic uncertainty. This balance between exploitation and exploration eventually will facilitate the search for global optima.

\subsection{Search Space Reduction}










While BayesOpt has some important benefits, its implementation often requires much computation to thoroughly search through the parameter space. The computational cost will be especially high when there are many parameters. The challenge carries over to the identification of the NDC-T model. To speed up the optimization process, we leverage the technique in~\cite{Perrone:Nips:2019} to reduce the search space. The key idea lies in using the existing evaluations and data points after every few iterations to determine the best search space, which is narrower in size, for the subsequent iterations.

We consider shaping the search space for $\boldsymbol{\theta}$ as an ellipsoid:
\begin{align*}
   \{\Theta\,|\,\left(\boldsymbol{\theta} - \boldsymbol{c}\right)^{\top} \boldsymbol{A} \left(\boldsymbol{\theta} - \boldsymbol{c}\right) \le 1 \},
\end{align*}
where  $\boldsymbol{c}$ is the center of the ellipsoid, and $\boldsymbol{A} > 0$ is the shape matrix. Suppose that we have collected a few data points in the foregoing search, and then pick the best $\tau$ data from them which have the largest $L$. They are denoted as $L(\boldsymbol{\theta}_{1:\tau})$ evaluated at $\boldsymbol{\theta}_\imath$ for $\imath=1,2,\ldots,\tau$. It is plausible to suppose that these $\tau$ data points approximately shape up a space that encompasses the maximum of $L(\boldsymbol{\theta})$.  We can determine the space by finding an ellipsoid such that $\boldsymbol{\boldsymbol{\theta}}_\imath$ for $\imath=1,2,\ldots,\tau$ will all lie in it. This is achieved through addressing the following optimization problem:
\begin{align*} 
   \nonumber
     \min_{\boldsymbol{A}, \boldsymbol{c}} \quad &\log \det(\boldsymbol{A}^{-1}), \\
      \textrm{s.t.} \quad
   &\left(\boldsymbol{\theta}_{\imath} - \boldsymbol{c}\right)^{\top} \boldsymbol{A} \left(\boldsymbol{\theta}_{\imath} - \boldsymbol{c}\right)  \le 1, \quad  \imath=1,2,\ldots,\tau, \\
   \nonumber
   & \boldsymbol{A}>0.
\end{align*}
The problem admits a solution known as the Khachiyan algorithm~\cite{Nima:Ellipsoid:2005}. We introduce rounds of BayesOpt, where one round consists of running BayesOpt for several iterations and then adjusting the parameter search space. By narrowing down the search space in this way, BayesOpt will gain more computational efficiency. The standard version of BayesOpt can be readily modified to incorporate a constrained search space, as shown in~\cite{Michael:AUAI:2014}.


To sum up, the advantages of using the proposed algorithm are three-fold: 
\begin{enumerate}
    \item The parameters are identified directly in the continuous-time space. Unlike other approaches, we avoid the burdensome procedure of identifying the discrete-time model first and then converting it back to the continuous-time model. The algorithm can also handle the use of any current load profiles.

    \item The algorithm is gradient-free. Therefore, there is no need to calculate the gradient with respect to $\boldsymbol{\theta}$.

    \item BayesOpt is provably effective at finding global optima~\cite{frazier:tutorial:2018}, presenting a promise to identify the NDC-T model accurately. Its computational demands are reasonable, especially when given the search space reduction. It ensures a physically meaningful estimate with reasonable computational time.
\end{enumerate}
Note that the proposed algorithm is applicable to a broader spectrum of ECMs and even electrochemical models, though we consider the NDC-T model specifically in this paper. 

\begin{table}[t!]\centering
\ra{1.2}
\caption{Initial parameter search space and comparison of the true and the identified parameters of the NDC-T model.}
 \begin{tabular}{ l | l l l }
\toprule
Parameters & Search range & True & Identified \\
\midrule
$C_b\ [\si{F}]$ & 7000$\sim$11000 & 10037 & 10043 \\
$C_s\ [\si{F}]$ & 700$\sim$1100 & 973 & 964 \\
$R_b\ [\si{\Omega}]$ & 0$\sim$0.1 & 0.019 & 0.0188 \\
$R_o\ [\si{\Omega}]$ & 0$\sim$0.1 & 0.026 & 0.0259 \\
$C_{\mathrm{core}}\ [\si{J/kelvin}]$ & 20$\sim$70 & 40 & 41.69 \\
$C_{\mathrm{surf}}\ [\si{J/kelvin}]$ & 0$\sim$20 & 10 & 13.67 \\
$R_{\mathrm{core}}\ [\si{kelvin/W}]$ & 0$\sim$10 & 4 & 2.80 \\
$R_{\mathrm{surf}}\ [\si{kelvin/W}]$ & 5$\sim$15 & 7 & 7.27 \\
$\kappa_1$ & 0$\sim$100 & 30 & 31.07 \\
$\kappa_2$ & 0$\sim$100 & 70 & 62.69 \\
\bottomrule
\end{tabular}
\label{Table: Result}
\end{table}

\begin{figure}[t!]
    \centering
    \includegraphics[width = 0.5\textwidth,trim={.5cm 1.2cm .6cm 1.3cm},clip]{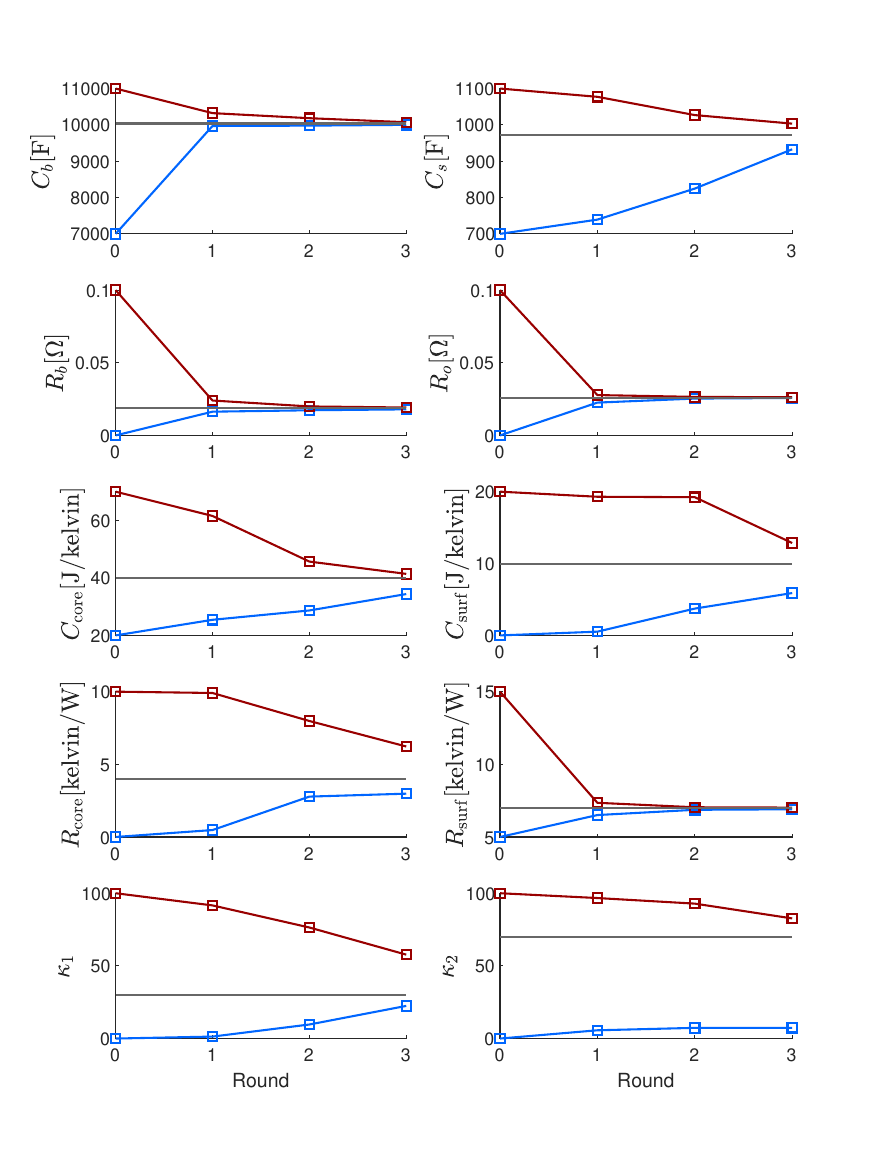}
    \caption{Upper and lower bounds on the search space for each parameter over four rounds of BayesOpt. ``$\color{darkred}-$" for upper bound, ``$\color{darkblue}-$" for lower bound, and ``$-$" for the true parameter.}
    \label{Fig:SpaceRedution}
\end{figure}

\section{Simulation Studies} \label{sec: Sim}

This section presents a simulation study to demonstrate the effectiveness of the proposed algorithm. 

We consider a 3.3 Ah NCA/graphite LiB cell, whose parameters are taken from~\cite{Tian:TCST:2021,Biju:AE:2023} and shown in Table~\ref{Table: Result}. Multiple synthetic datasets are generated by discharging the cell from full ($\mathrm{SoC}(0)=1$) using the US06, UDDS and LA92 current profiles~\cite{EPA}. Each current profile is scaled to be between 0$\sim$4 A. The datasets also account for different ambient temperatures so as to identify the thermal parameters. The $T_\mathrm{amb}$ is set to be 313 kelvin, 283 kelvin, and 298 kelvin for the US06, UDDS, and LA92 profiles, respectively. The sampling interval is 1~\si{s}. The covariances of the measurement noises are set to be $R_{V} = 10^{-4}$ and $R_T=10^{-3}$. The reference temperature is $T_{\mathrm{ref}}=298$ kelvin. In the simulations, the cell's initial state is known, which is $V_b(0)=V_s(0)=1$ and $T_c(0)=T_s(0)=T_{\mathrm{amb}}$. 

\begin{figure}[t!]
    \centering
    \includegraphics[width = 0.49\textwidth,trim={.25cm 0cm .2cm 1cm},clip]{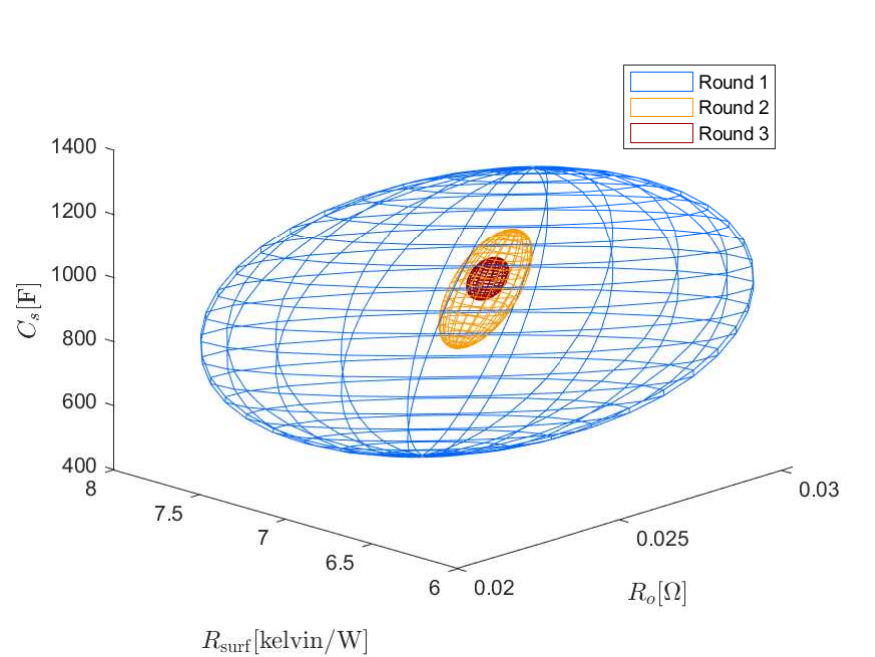}
    \caption{Projection of the ellipsoid-shaped parameter search space on $C_s$, $R_\mathrm{surf}$ and $R_o$ in round one to three of BayesOpt.}
    \label{Fig:Ellipsoid}
\end{figure}

\begin{figure}[t!]
    \centering
    \includegraphics[width = 0.5\textwidth,trim={.5cm .2cm .8cm .3cm},clip]{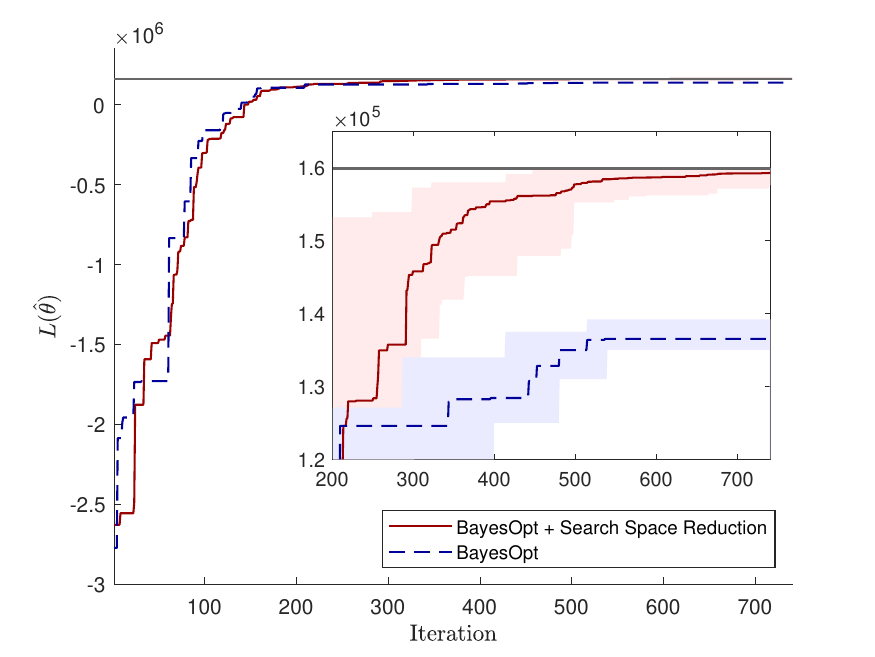}
    \caption{Average maximum log-likelihood over 20 independent runs. $``-"$ shows the log-likelihood evaluated at the true parameters.}
    \label{Fig:Likelihood}
\end{figure}

In implementing BayesOpt, we use the initial parameter search space shown in Table~\ref{Table: Result}, and narrow the search space three times, after every $200$ iterations. The ellipsoidal search space is determined using the best $20$ data points. Because the parameters vary across different orders of magnitude, they are pre-processed by normalization before feeding to the Gaussian process, as often recommended in the practice of Gaussian process regression. 

\begin{figure}[t!]
    \centering
    \includegraphics[width = 0.5\textwidth,trim={.8cm .9cm 1cm .9cm},clip]{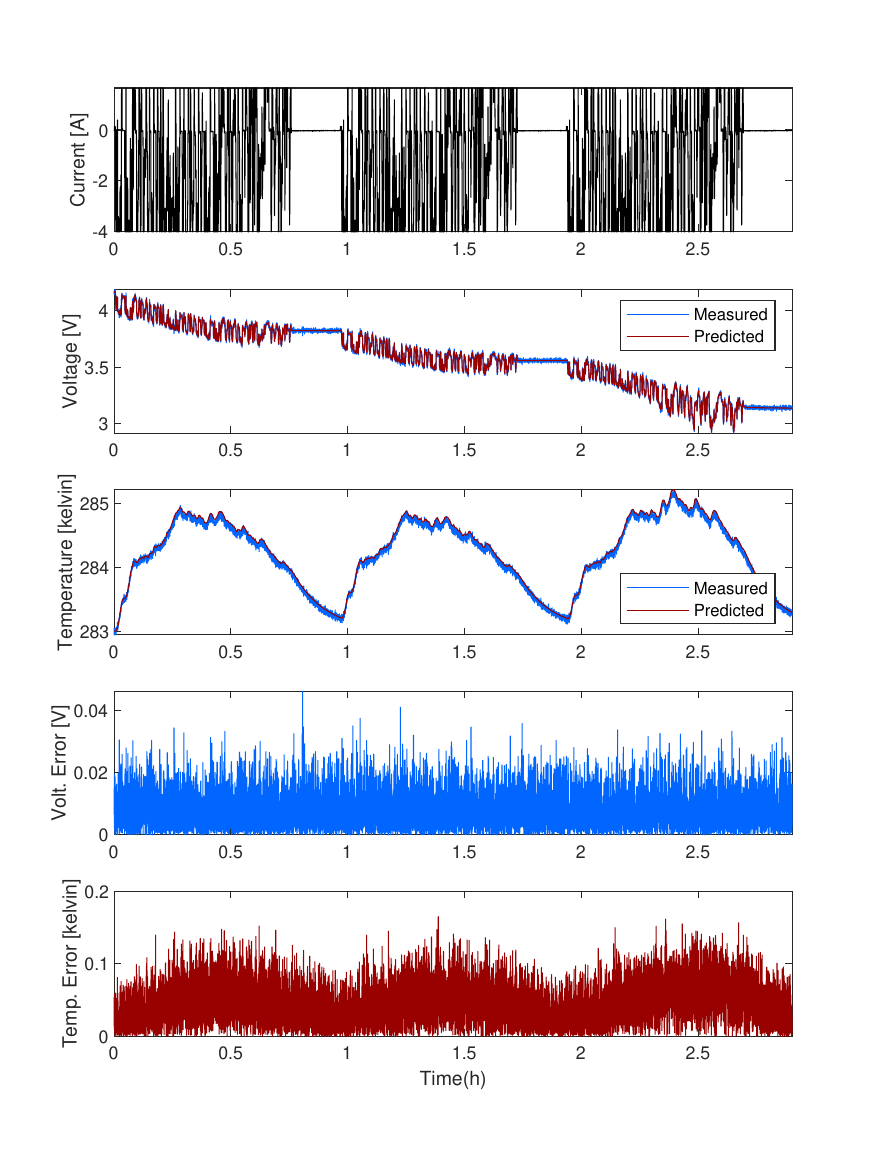}
    \caption{Comparison of the measured and predicted voltage and surface temperature under the UDDS discharging profile at $T_{\mathrm{amb}}=283$ kelvin.}
    \label{Fig:CompResult}
\end{figure}

The parameter estimation is shown in Table~\ref{Table: Result}, which well agrees with the truth. The figures are further supplementary evidence. Figs.~\ref{Fig:SpaceRedution}-\ref{Fig:Ellipsoid} show that the search space reduction in the optimization process. The shrinking size implies that the parameter estimation performance improves. Fig.~\ref{Fig:Likelihood} shows the $L(\hat{\boldsymbol{\theta}})$ with respect to the iterations. The comparison with the standard BayesOpt also shows that the search space shrinking accelerates the search.  Fig.~\ref{Fig:CompResult} illustrates a comparison between the predicted voltage and surface temperature and the measurements when the battery is applied with the UDDS current profile. Throughout the discharging process, the voltage error is within 0.04 V, and the temperature error is within 0.2 kelvin. These results indicate that the proposed algorithm delivers good system identification performance.

\section{Conclusions} \label{sec: Conclusion}
System identification for LiBs has attracted perennial interest due to its essential role in various applications. The problem becomes challenging and intriguing when a LiB model integrates the electrical and thermal dynamics to be nonlinear and complex. We consider such an equivalent circuit model in this study, the NDC-T model, and harnesses BayesOpt to enable the maximum likelihood estimation of its parameters. The use of BayesOpt is motivated by its power in globally optimizing hard-to-evaluate objective functions through probabilistic machine learning. To speed up the implementation of BayesOpt for the considered problem, we add a procedure to reduce the parameter search space dynamically. The proposed parameter identification approach shows some useful merits for practical LiB applications. First, it can well handle the NDC-T model which has a nonlinear continuous-time state-space representation. Second, it extracts the physical parameters directly from measurement data. Finally, it allows the use of a wide range of current profiles. We validate the effectiveness of the proposed method through simulation and highlight its potential for identifying more equivalent circuit models. 










\bibliographystyle{ieeetr}
\bibliography{ref}

\end{document}